# Direct observation of dynamic modes excited in a magnetic insulator by pure spin current

V. E. Demidov[1*], M. Evelt[1], V. Bessonov[2], S. O. Demokritov[1,2], J. L. Prieto[3], M. Muñoz[4], J. Ben Youssef[5], V. V. Naletov[6,7], G. de Loubens[6], O. Klein[8], M. Collet[9], P. Bortolotti[9], V. Cros[9] and A. Anane[9]

[1]*Institute for Applied Physics and Center for Nanotechnology, University of Muenster, 48149 Muenster, Germany*

[2]*M.N. Miheev Institute of Metal Physics of Ural Branch of Russian Academy of Sciences, Yekaterinburg 620041, Russia.*

[3]*Instituto de Sistemas Optoelectrónicos y Microtecnologa (UPM), Ciudad Universitaria, Madrid 28040, Spain*

[4]*IMM-Instituto de Microelectrónica de Madrid (CNM-CSIC), PTM, E-28760 Tres Cantos, Madrid, Spain*

[5]*Laboratoire de Magnétisme de Bretagne CNRS, Université de Bretagne Occidentale, 29285 Brest, France*

[6]*Service de Physique de l' État Condensé, CEA, CNRS, Université Paris-Saclay, CEA Saclay, 91191 Gif-sur-Yvette, France*

[7]*Institute of Physics, Kazan Federal University, Kazan 420008, Russian Federation*

[8]*INAC-SPINTEC, CEA/CNRS and Univ. Grenoble Alpes, 38000 Grenoble, France*

[9]*Unité Mixte de Physique CNRS, Thales, Univ. Paris Sud, Université Paris-Saclay, 91767 Palaiseau, France*

*Abstract:*

Excitation of magnetization dynamics by pure spin currents has been recently recognized as an enabling mechanism for spintronics and magnonics, which allows implementation of spin-torque devices based on low-damping insulating magnetic materials. Here we report the first spatially-resolved study of the dynamic modes excited by pure spin current in nanometer-thick microscopic insulating Yttrium Iron Garnet disks. We show that these modes exhibit nonlinear self-broadening preventing the formation of the self-localized magnetic bullet, which plays a crucial role in the stabilization of the single-mode magnetization oscillations in all-metallic systems. This peculiarity associated with the efficient nonlinear mode coupling in low-damping materials can be among the main factors governing the interaction of pure spin currents with the dynamic magnetization in high-quality magnetic insulators.

One of the decisive advantages of pure spin currents for the emerging technologies, such as spintronics[1-3] and magnonics[4-6], is the possibility to excite the magnetization dynamics in insulating magnetic materials by the spin-transfer torque[7,8]. The breakthrough idea that the spin-transfer torque can be utilized to induce spin waves in magnetic insulators suggested in the pioneering work by Kajiwara *et al.* (Ref. 7) had an enormous impact on the developments in the physics of magnetism and revived interest to magnetic insulators, such as Yttrium Iron Garnet (YIG). The main advantage of this material is the unmatched small dynamic magnetic damping, which is expected to result in more efficient operation of spin-torque devices by enabling significant reduction of the density of the driving current necessary for the onset of current-induced auto-oscillations. Although this advantage should greatly simplify the implementation of YIG-based spin-torque devices, it took many years of intense research to achieve the spin-current induced excitation of coherent magnetization dynamics in this material[9-10].

In contrast, significantly more rapid progress has been achieved in the studies of all-metallic spin-current driven systems, which has led in the recent years to the demonstration of a large variety of nano-devices exhibiting highly coherent magnetic oscillations driven by the pure spin current[11-18]. This progress was greatly facilitated by the flexible geometry of spin-current driven systems enabling the direct magneto-optical imaging of the current-induced dynamics. In particular, the spatially-resolved measurements have shown that the significant role in the stabilization of the single-mode current-induced oscillations is played by the nonlinear spatial self-localization effects resulting in the formation of the so-called bullet mode[11,19], which is a two-dimensional standing analogue of a spin-wave soliton[20].

Since the dynamic nonlinearity of the spin system largely determines the behaviors of the dynamic magnetization under the influence of the spin current, the peculiarities of the nonlinear response of low-damping YIG can be among the most important factors affecting the current-induced excitation of the magnetization dynamics in this material. In fact, it is long-known that the low damping in YIG facilitates the nonlinear coupling between dynamic magnetic modes[21]. This phenomenon was found to strongly affect the formation of spin-wave solitons[22] and result in a suppression of the nonlinear spatial self-localization phenomena[23], which play a crucial role in spin-torque driven nano-systems[17].

Here we study experimentally the effects of the dynamic nonlinearity on the spatial characteristics of dynamic modes excited by the pure spin current in microscopic YIG disks. By using the direct spatially-resolved imaging of the modes, we show that the dynamic magnetization exhibits strong spatial localization at the onset of the auto-oscillations. However, just above the onset, the amplitude of the auto-oscillations rapidly saturates and the excited mode experiences a spatial broadening. This phenomenon is opposite to the nonlinear self-localization observed in all-metallic systems and can be the main factor determining the response of the magnetization in low-damping materials to the spin current. We attribute these behaviors to the efficient nonlinear mode coupling, which prevents the growth of the amplitudes of the dynamic magnetization to the level necessary for the onset of nonlinear self-localization and formation of the bullet mode.

**Results**

**Test devices.** The schematic of our experiment is shown in Fig. 1. The studied device consists of a YIG(20 nm)/Pt(8 nm) disk with a 2 μm diameter (details about the preparation of the YIG films can be found in Ref. 24). The Pt layer is electrically contacted by using two Au(80 nm) electrodes with a 1 μm wide gap between them. By applying an electric voltage between the electrodes, we inject the dc electric current into the Pt layer. Because of the large difference in the electric conductance of the electrodes and the Pt film, the current is injected into the Pt layer only in the electrode gap area. Due to the spin-Hall effect[25,26], the in-plane electrical current in the Pt film is converted into a transverse spin accumulation. The associated pure spin current is flowing in the out-of-plane direction and is injected into the YIG film[7] resulting in a spin-transfer torque (STT)[27,28] on its magnetization $M$. The device is magnetized by the in-plane static magnetic field $H_0$ applied perpendicular to the direction of the dc current flow. For positive currents, as defined in Fig. 1, the polarization of the spin current corresponds to the STT compensating the dynamic damping in the magnetic film. For sufficiently large dc currents, the damping becomes completely compensated, which results in the excitation of magnetic auto-oscillations[10].

**Magnetooptical measurements.** To observe the oscillations induced by the spin current we use micro-focus Brillouin light scattering (BLS)[29] spectroscopy, which enables the spectrally- and the spatially-resolved measurements of the dynamic magnetization. We focus probing laser light into a diffraction-limited spot on the surface of YIG (Fig. 1) and analyze the spectrum of light inelastically scattered from the magnetization oscillations. The resulting BLS signal at the selected frequency is proportional to the intensity of the dynamic magnetization at this frequency, at the location of the laser spot. The wavelength of the laser is chosen to be 473 nm, which provides high sensitivity of the method for measurements with ultra-thin YIG. The power of the probing light is as low as 0.05 mW, which guarantees negligible laser-induced heating of the sample.

In Fig. 2a we show BLS spectra recorded by placing the probing laser spot in the middle of the YIG disk for different values of the dc current $I$. The spectrum shown by solid black symbols recorded for $I = 0$ characterizes the magnetic excitations existing at room temperature due to the thermal fluctuations. Similar to the all-metallic systems[11], the application of the dc current results in the gradual enhancement of the magnetic fluctuations followed at $I = I_C \approx 8$ mA by an abrupt emergence of a narrow intense peak (blue squares in Fig. 2a) marking the onset of the current-induced auto-oscillations of the YIG magnetization. The intensity of the auto-oscillation peak exceeds that of the thermal fluctuations at $I = 0$ by about two orders of magnitude. Further increase in $I$ results in the lowering of the central frequency of the peak and in its spectral broadening (red diamonds in Fig. 2a).

To obtain information about the dynamical modes contributing to the current-induced auto-oscillations, we perform two-dimensional mapping of the dynamic magnetization in the YIG disk by rastering the probing laser spot in the two lateral directions. In Fig. 2b we show a typical spatial map obtained for $I = I_C+1$ mA. As seen from these data, the auto-oscillations are strongly localized in the $y$-direction in the area corresponding to the gap between the electrodes, whose edges are marked in Fig. 2b by the horizontal lines. This localization can be attributed to the formation of an effective confining potential due to the joint effects of the spin- and electrical currents, as discussed below. In contrast, there is no pronounced localization of the auto-oscillations in the $x$-direction. They clearly extend over the entire active device area, where the injection of the spin current takes place.

Figure 3 illustrates the spatial localization of the auto-oscillations for different values of the dc current. $x$-profiles of the oscillation intensity recorded at different $I$ (Fig. 3a) show that the auto-oscillations do not always occupy the entire active device area. At currents close to $I_C$, they are strongly localized in the middle of the gap. However, with the increase in $I$, they rapidly expand in the $x$ direction. This process is characterized in detail in Fig. 3b, which shows the current dependence of the full width at half maximum (FWHM) of the $x$-profiles. This width linearly increases from 0.5 μm to about 1.4 μm for $I$ varying from $I_C$ to $I_C$+1 mA and then saturates at the value of 1.5 μm.

We emphasize that these behaviors are in contradiction with the typical manifestations of the dynamic magnetic nonlinearity, which is expected to result in the spatial self-localization of the intense oscillations and lead to a formation of the spin-wave bullet[19] experimentally observed in all-metallic systems driven by the pure spin current[11,17]. The absence of the self-localization in YIG can be attributed to a nonlinear limitation of the amplitude of the auto-oscillations. Indeed, if this amplitude is limited by any other nonlinear phenomenon, the self-localization mechanism requiring the dynamic magnetization to be comparable with the static magnetization can be suppressed.

This scenario is supported by the data of Fig. 3c. The maximum intensity of the auto-oscillations (up triangles in Fig. 3c) saturates immediately after the onset and then stays nearly constant over the entire used range of $I$. In contrast, the integral intensity (down triangles in Fig. 3c) exhibits a gradual increase up to $I$ = 9.2 mA. This indicates that the amplitude of the dynamic mode excited at the onset cannot grow above a certain level and that the further increase in $I$ results in the energy flow into other dynamical modes leading to the formation of a strongly nonlinear spatially-extended mode, which cannot be treated as a linear combination of the eigenmodes of the system[23]. These behaviors can be attributed to the efficient nonlinear mode coupling stimulated by the small dynamic damping in YIG. We emphasize that the same mechanism was previously found to result in the formation of traveling dark solitons in a medium characterized by the attractive nonlinearity, where the self-localization effects are expected to lead to the formation of bright solitons[22].

According to the described scenario, the current-induced auto-oscillations are also expected to demonstrate a spatial broadening in the *y*-direction. The corresponding experimental data are shown in Fig. 4. We note that the *y*-profiles of the auto-oscillations (Fig. 4a) are mostly determined by the confining potential, which prevents the extension of the dynamic excitations outside the region of the gap between the electrodes. Because of the limited spatial resolution of the used technique, which can be estimated as 240-250 nm, one cannot quantitatively analyze the broadening of the *y*-profiles with the increase in the dc current. However, from Fig. 4b it is seen, that the *y*-width also tends to increase with increasing *I*.

**Discussion**

Finally, we address the nature of the localized dynamic mode, which is excited in the YIG disk at $I = I_C$ before the nonlinear broadening takes place. To understand why this mode shows a strong localization at the center of the disk, we perform ferromagnetic-resonance (FMR) measurements based on the excitation of the system by a dynamic magnetic field created by an additional 5 μm wide stripe antenna aligned parallel to the gap between the electrodes (see also Supplementary Figure 1). In Figs. 5a and 5b we show the spatial maps of the field-driven FMR mode recorded by BLS without any dc current in Pt and at $I = 7$ mA $< I_C$, respectively. The maps clearly demonstrate that the FMR mode initially spread over the entire disk becomes localized at its center when the dc current is applied. We emphasize that the measurements were performed at sufficiently small microwave powers to exclude any nonlinear self-localization effects[30]. Therefore, the observed localization can only be attributed to the joint effect of the Oersted field of the dc current, the local reduction of the magnetization due to the Joule heating, and the partial compensation of the dynamic damping by the pure spin current injected into the YIG film.

To prove this assumption we perform micromagnetic simulations using the software package MuMax3 (Ref. 31) (see Methods for details). The obtained spatial maps of the intensity of the dynamic magnetization in the FMR mode are shown in Figs. 5c and 5d for the cases of $I$ = 0 and 7 mA, respectively. As seen from the comparison with the experimental data (Fig. 5a and 5b), the simulations reproduce well the localization behavior and clearly show that the effects of the driving current result in a shrinking of the FMR mode in both lateral directions resulting in the formation of a strongly localized linear mode of the system. It is this mode that is most probably first excited by the pure spin at the onset of the auto-oscillations (see corresponding profiles in Figs. 3a and 4a). It is to be noticed that these current-induced localization effects counteract the nonlinear broadening. Therefore, in a properly designed system, they can be utilized to suppress this detrimental phenomenon.

In conclusion, we demonstrate that the dynamic nonlinearity of the low-damping YIG films leads to a nonlinear self-broadening of the current-induced oscillations instead of the nonlinear self-localization typical for metallic ferromagnetic films characterized by a relatively large dynamic damping. Since the self-localization mechanisms play a crucial role for the stabilization of the single-mode current-induced auto-oscillations, in order to achieve highly-coherent oscillations in the YIG films, it is necessary to utilize additional confining mechanisms, such as the patterning of the film into a nano-size element or creation of an additional confining potential. Our results shine light on the physical mechanisms responsible for the peculiarities of the interaction of magnetization in YIG with pure spin currents and create a base for implementation of efficient insulator-based spin-torque nano-devices.

**Methods**

**Sample fabrication.** 20 nm thick YIG films were grown by the pulsed laser deposition on Gadolinium Gallium Garnet (GGG) (111) substrates[24]. The 8 nm thick layer of Pt was deposited using dc magnetron sputtering. The dynamical properties of bare YIG films and YIG/Pt bilayers have been determined from broadband FMR measurements. The YIG/Pt microdiscs and the Au(80 nm)/Ti(20 nm) electrodes were defined by e-beam lithography. The system was insulated by a 300 nm thick $SiO_2$ layer, and a broadband 5 μm wide microwave antenna made of 250 nm thick Au was defined on top of the system using optical lithography.

**Magneto-optical measurements.** Micro-focus BLS measurements were performed by focusing light produced by a continuous-wave single-frequency laser operating at the wavelength of 473 nm into a diffraction-limited spot. The light scattered from magnetic oscillations was analyzed by a six-pass Fabry-Perot interferometer TFP-1 (JRS Scientific Instruments, Switzerland) to obtain information about the BLS intensity proportional to the square of the amplitude of the dynamic magnetization at the location of the probing spot. By rastering the spot over the surface of the sample using a closed-loop piezo-scanner, two-dimensional maps of the dynamic magnetization were recorded with the spatial step size of 100 nm. The positioning system was stabilized by custom-designed software-controlled active feedback, providing long-term spatial stability of better than 50 nm. All measurements were performed at room temperature.

**Simulations.** The micromagnetic simulations were performed by using the software package MuMax3 (Ref. 31). The computational domain with dimensions of $2\times2\times0.02$ μm$^3$ was discretized into $10\times10\times10$ nm$^3$ cells. The FMR mode was excited by a 50 ps wide pulse of the out-of-plane magnetic field. The spatial map of the dynamic magnetization was reconstructed based on the Fourier analysis of the dynamic response of the magnetization. Standard value for the exchange stiffness of $3.66\times10^{-12}$ J/m was used, while the value of the saturation magnetization $4\pi M_0$=2.2 kG and the Gilbert damping constant of $2\times10^{-3}$ were determined from the FMR measurements. For the case of the disk free from the influence of the driving current, the saturation magnetization $M$, the damping constant $\alpha$, and the static field $H_0$=1000 Oe were considered to be uniform across the disk area. To take into account the effects of the driving current, the parameters $M$, $\alpha$, and $H_0$ were reduced in the central rectangular region corresponding to the area of the gap between the electrodes. The reduction of $M$ was taken from the results of the FMR measurements (Supplementary Figure 1), the reduction of $\alpha$ was calculated assuming the linear dependence of effective damping on the current strength, and the reduction of the field was calculated by using the Ampere's law.

**\* Corresponding author:** Correspondence and requests for materials should be addressed to V.E.D. (demidov@uni-muenster.de).

**Acknowledgements:** We acknowledge E. Jacquet, R. Lebourgeois, R. Bernard and A. H. Molpeceres for their contribution to sample growth, and O. d'Allivy Kelly and A. Fert for fruitful discussion. This research was partially supported by the Deutsche Forschungsgemeinschaft, the ANR Grant Trinidad (ASTRID 2012 program), and the program Megagrant № 14.Z50.31.0025 of the Russian Ministry of Education and Science. M.C. acknowledges DGA for financial support. V. V. N. acknowledges support from the Competitive Growth of KFU.

**Author Contributions:** VED, ME and VB performed BLS measurements and data analysis. ME additionally performed micromagnetic simulations. AA and VC supervised the growth of the YIG films. JBY performed the magnetic characterizations. MC, VVN, JLP, MM, GdL and AA designed, nanofabricated, and characterized the samples. VVN, PB, JBY, VED, SOD, VC, AA, GdL and OK initiated and conducted the project. All authors co-wrote and discussed the manuscript.

**Additional Information:** The authors have no competing financial interests.

**Figure legends**

**Figure 1. Experimental layout.** YIG(20 nm)/Pt(8 nm) disk with the diameter of 2 μm is electrically contacted by using two Au(80 nm) electrodes with a 1 μm wide gap between them. The in-plane dc electrical current in the Pt film is converted into the out-of-plane spin current by the spin-Hall effect. The spin current is injected into the YIG film and exerts the spin-transfer torque on its magnetization *M* resulting in the excitation of magnetic auto-oscillations. The excited oscillations are detected by the magneto-optical technique utilizing probing laser light focused through the sample substrate.

**Figure 2. Spectral and spatial characteristics of the auto-oscillations. a,** BLS spectra recorded by placing the probing laser spot in the middle of the YIG disk for different strength of the dc current *I*: black solid symbols – $I = 0$, blue open squares – $I = I_C \approx 8$ mA, red open diamonds – $I = I_C+1$ mA. The intensity of the peak for $I = I_C$ is by about two orders of magnitude larger compared to that for $I = 0$. Lines are the guides for the eye. The spectral linewidth of the peaks is determined by the limited spectral resolution of the measurement apparatus. The data were recorded at $H_0 = 1000$ Oe. **b,** Typical spatial map of the intensity of current-induced magnetic auto-oscillations in the YIG disk. The contours of the disk and the edges of the electrodes are shown by white lines. The map was recorded for $I = I_C+1$ mA and $H_0 = 1000$ Oe.

**Figure 3. Dependence of the spatial localization of the auto-oscillations on the dc current. a,** Spatial profiles of the oscillation intensity in the direction along the gap between the electrodes recorded for different dc currents, as labeled. **b,** Current

dependence of the full width at half maximum (FWHM) of the spatial profiles. **c,** Current dependences of the maximum intensity of the auto-oscillations detected in the center of the gap and that of the spatially-integrated intensity. The data were recorded at $H_0$ = 1000 Oe. Symbols are experimental data, curves are guides for the eye.

**Figure 4. Spatial localization of the auto-oscillations across the gap between the electrodes. a,** Spatial profiles of the oscillation intensity recorded for different dc currents, as labeled. **b,** Current dependence of the full width at half maximum (FWHM) of the spatial profiles. The data were recorded at $H_0$ = 1000 Oe. Symbols are experimental data, curves are guides for the eye.

**Figure 5. Spatial characteristics of the field-driven ferromagnetic resonance mode. a,** and **b,** Spatial maps of the FMR mode measured by BLS at $I$ = 0 and 7 mA, respectively. The contours of the disk and the edges of the electrodes are shown by white lines. The maps were recorded at $H_0$ = 1000 Oe. **c,** and **d,** Normalized spatial maps of the intensity of the dynamic magnetization in the FMR mode obtained from micromagnetic simulations. **c** corresponds to $I$ = 0 and **d** corresponds to $I$ = 7 mA.

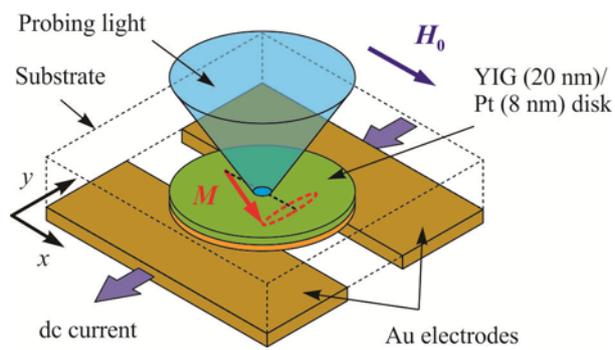

Fig. 1

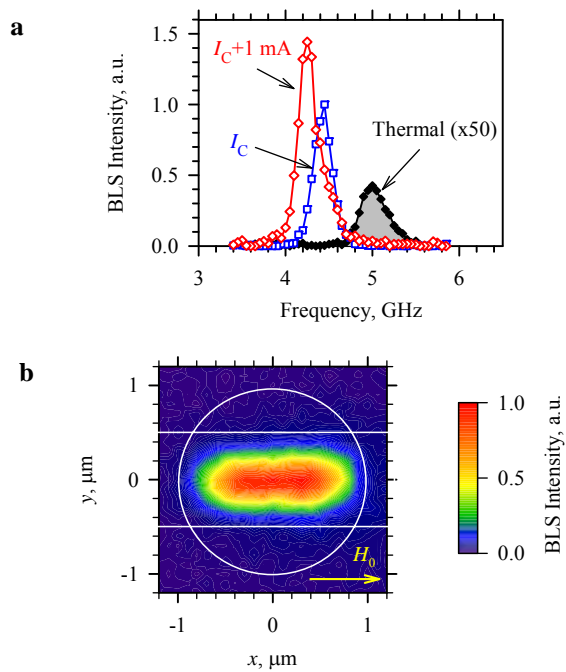

Fig. 2

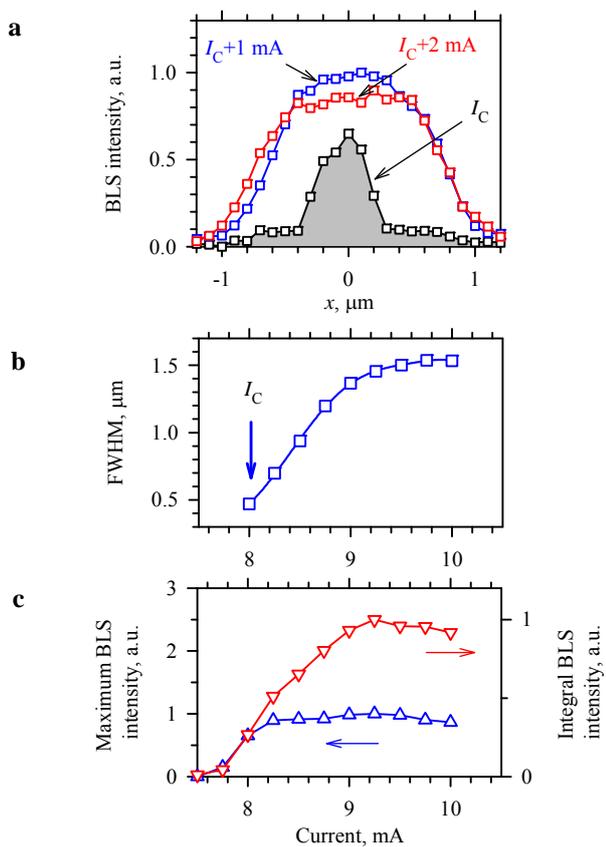

Fig. 3

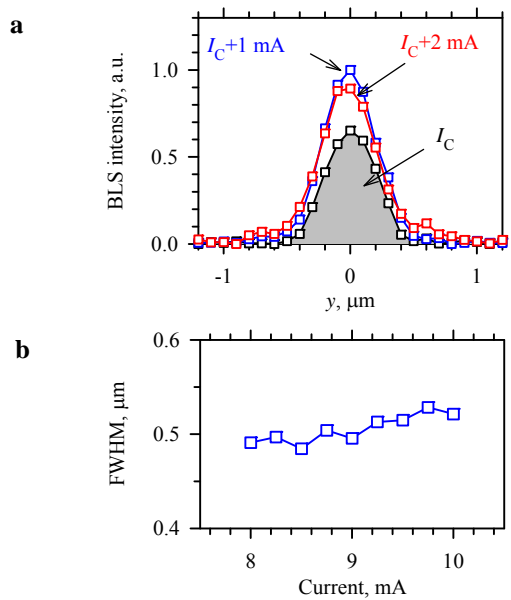

Fig. 4

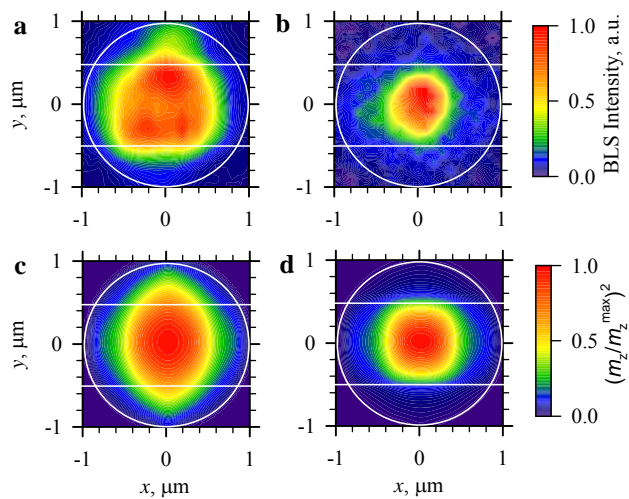

Fig. 5

**Supplementary figures.**

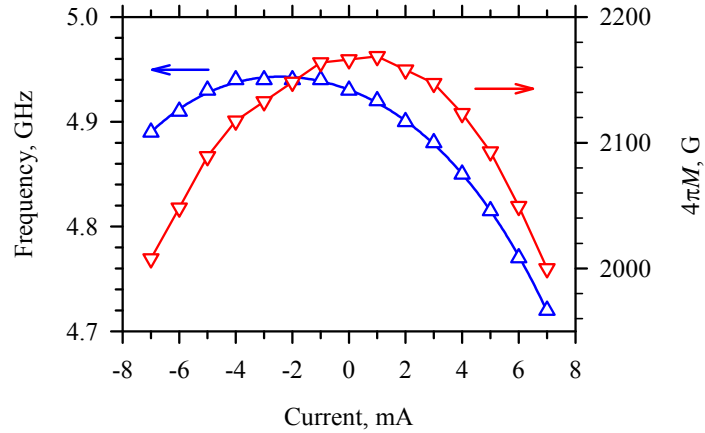

**Supplementary Figure 1. BLS measurements of the ferromagnetic resonance in the YIG disk.** In order to measure the ferromagnetic resonance (FMR) in the YIG disk, we transmit a microwave current through an additional 5 μm wide stripe antenna, which creates a nearly uniform in-plane dynamic magnetic field across the YIG disk. By varying the frequency of the excitation signal in the antenna and measuring the BLS intensity, we record the FMR curves and obtain the FMR frequency. Additionally, the electrical dc current $I<I_C$ is applied through the Pt layer to analyze its influence on the FMR. The FMR frequency obtained for different currents is shown by up-triangles. The observed variation of the frequency originates from the Oersted field of the current and the reduction of the static magnetization $M$ caused by the Joule heating of the sample. We estimate the Oersted field from the Ampere's law and calculate the current dependence of the magnetization by using the Kittel formula for the FMR frequency. The results shown by the down-triangles indicate that $M$ is reduced by about 8% at $I=7$ mA. The obtained dependence is nearly symmetrical with respect to the change of the sign of the current, in agreement with the expectations for the effects of the Joule heating. The data were obtained at $H_0=1000$ Oe. Lines are guides for the eye.